# Phases and magnetism at the microscale in compounds containing nominal $Pb_{10-x}Cu_x(PO_4)_6O$

Chang Liu[1,†], Wenxin Cheng[1,2,†], Xiaoxiao Zhang[1,2,†], Juan Xu[1,3], Jiaxin Li[1], Qiuyan Shi[1,2], Changhong Yuan[1,2], Li Xu[1,2], Honglin Zhou[1,2], Shilin Zhu[1], Jianping Sun[1], Wei Wu[1], Jianlin Luo[1], Kui Jin[1,2,4], Yangmu Li[1,2,*]

[1]Beijing National Laboratory for Condensed Matter Physics, Institute of Physics, Chinese Academy of Sciences, Beijing 100190, China
[2]School of Physical Sciences, University of Chinese Academy of Sciences, Beijing 100049, China
[3]School of Physics and Technology, Wuhan University, Wuhan 430072, China
[4]Songshan Lake Materials Laboratory, Dongguan, Guangdong 523808, China

†: These authors contributed equally to this work
*: Corresponding author: yangmuli@iphy.ac.cn

**Abstract:**

Achieving superconductivity at room temperature could lead to substantial advancements in industry and technology. Recently, a compound known as Cu-doped lead-apatite, $Pb_{10-x}Cu_x(PO_4)_6O$ (0.9 < x < 1.1), referred to as "LK-99", has been reported to exhibit unusual electrical and magnetic behaviors that appear to resemble a superconducting transition above room temperature. In this work we collected multiphase samples containing the nominal $Pb_{10-x}Cu_x(PO_4)_6O$ phase (no superconductivity observed in our measured samples), synthesized by three independent groups, and studied their chemical, magnetic, and electrical properties at the microscale to overcome difficulties in bulk measurements. Through the utilization of optical, scanning electron, atomic force, and scanning diamond nitrogen-vacancy microscopy techniques, we are able to establish a link between local magnetic properties and specific microscale chemical phases. Our findings indicate that while the $Pb_{10-x}Cu_x(PO_4)_6O$ phase seems to have a mixed magnetism contribution, a significant fraction of the diamagnetic response can be attributed to Cu-rich regions (e.g., $Cu_2S$ derived from a reagent used in the synthesis). Additionally, our electrical measurements reveal the phenomenon of current path switch and a change in resistance states of $Cu_2S$. This provides a potential explanation for the electrical behavior observed in compounds related to $Pb_{10-x}Cu_x(PO_4)_6O$.

Superconductivity has been one of the most attractive areas of research since its discovery in 1911 [1]. Its two fundamental properties - zero electrical resistance and perfect diamagnetism - make it immensely valuable in various applications. Researchers have made great efforts to explore new superconductors, with the aim of achieving higher and higher transition temperatures [e.g., 2-6]. Among them, the cuprate superconductors, discovered in 1986 [2], currently hold the record for the highest superconducting transition temperature under ambient pressure [7]. In contrast, there also exist many instances of "unidentified superconducting objects" (USOs) that initially appeared promising but eventually proved otherwise [8].

Recently, Lee *et al.* studied a Cu-doped lead-apatite compound $Pb_{10-x}Cu_x(PO_4)_6O$ ($0.9 < x < 1.1$) (also known as "LK-99") and found highly unusual electrical and magnetic behaviors occurring above room temperature. Additionally, they reported a (partial) levitational effect of the material above permanent magnets [9,10]. These observations have kindled intense interest within scientific communities and beyond, owing to the low materials cost and potential straightforward synthesis of this compound. However, materials produced using the published LK-99 synthesis recipe have yielded vastly disparate outcomes and raised numerous questions, and the report of room-temperature superconductivity has not been firmly reproduced (e.g., [11–22]). One notable work by Zhu *et al.* reported that single-phase $Cu_2S$, a frequently encountered minor phase in currently available LK-99- related samples, undergoes a structural transition around 380 K, which could explain some of the unusual electrical properties of LK-99 [18]. Another work by Puphal *et al.* studied crystalline $Pb_{10-x}Cu_x(PO_4)_6O$ material and observed a diamagnetic response characteristic as well as a small ferromagnetic component [22]. These findings call for a closer examination of this material at the microscale, since bulk measurements often provide volume-averaged information.

In the present study, we collected compounds synthesized by three independent groups (denoted as C, L, and W samples), and studied their chemical, magnetic, and electrical properties at the microscale. The samples were grown mostly following the LK-99 synthesis steps [9,10]. Synthesis details for the W set of samples are reported in ref. [18], and very similar protocols were used for C and L samples. For W samples, a mixture of Cu and P powder with the molar ratio of 3:1 was sealed in high vacuum and heated to obtain $Cu_3P$ (300 °C at a rate of 4 °C/min, dwelled for 20 hours, then raised to 900 °C in 10 hours, followed by 10 additional hours of dwell). A stoichiometric mixture of $PbSO_4$ and $PbO$ powders was used for $Pb_2(SO_4)O$ (heated to 725°C at 4°C/min, then dwelled for 24 hours). Differences in molar ratio between $Cu_3P$ and Lanarkite $Pb_2(SO_4)O$ powders, which were used for the solid-state reaction by different groups, resulted in a variance of the microscale phases within the final samples. For C and W, the molar ratio of $Cu_3P$ and $Pb_2(SO_4)O$ used was 1:1, while for L it was 5:3. For W, the powders were pressed into pellets and sealed into evacuated fused silica ampoules (heated to 925°C at 5°C/min, dwelled for 24 hours, then annealed again at 925°C for 24 hours after cooling, regrinding, and repressing). We label all sample pieces with C, L, and W followed by identifying numbers. Sample information is summarized in Table 1.

We begin by presenting optical microscope images, captured using a Nikon ECLIPSE LV100ND, of representative C, L, and W samples in Fig. 1. 10,000-grit sandpapers were used to polish sample surfaces when necessary. Micro-phase separation is evident in all the measured samples. For simplicity, we will refer to the $Pb_{10-x}Cu_x(PO_4)_6O$ related phase as the "Pb phase" in the subsequent text. The Pb phase dominates the volume fraction of sample C, while Cu-related phases (i.e., $Cu_2S$, Cu, $CuO_x$) make up the majority of L and W. In the optical images, the gold-colored, slightly gray, and dark gray areas mainly correspond to Cu, $Cu_2S$, and Pb phases, respectively (see Figs. 2 and 3 for identification details). These microphases constitute islands and mazes interconnected in a three-dimensional manner within the samples, thereby complicating the electrical and magnetic responses.

Remarkably, $Cu_2S$, one of the secondary phases in LK-99-related compounds, has been found to be possibly responsible for the superconductivity-like electrical behavior exhibited by LK-99 [18]. To probe the impact of $Cu_2S$, we conducted post-growth annealing on C and L samples, and subsequently compared their characteristics before and after this process. Figure 1(f-j) shows zoomed-in optical microscope images of the as-grown/annealed C, as-grown L, and as-grown W samples. The annealing process induced a clear oxidation of $Cu_2S$/Cu phases (refer to chemical analysis in Fig. 3). To our surprise, a non-negligible portion of $Cu_2S$ remains in annealed samples and in W. To achieve a complete oxidation of the $Cu_2S$/Cu phases might require an extension of the annealing duration.

Figure 1(k-o) displays scanning electron microscopy (SEM) back scattering electron (BSE) images of the same samples characterized by optical microscopy, using a Hitachi SU5000 system with IXRF energy dispersive X-ray spectroscopy (EDS) capability. The red boxes in the SEM images highlight areas analyzed to obtain EDS spectra in Fig. 3. The black-white contrast of BSE images is associated with the atomic number of the chemical elements distributed within the microphases. White regions indicate the Pb phase, while darker gray areas denote the Cu-related phases.

In Figure 2(a,b), we plot x-ray powder diffraction patterns of as-grown C and W (data for W are taken from ref. [18]). For both samples, minor phases such as $Cu_2S$ and Cu are revealed to coexist with the Pb phase. Figure 2(c) demonstrates how we crossmatch between optical and SEM images, which serves as the basis for determining phase and chemical distributions. Corresponding EDS data from the box areas depicted in Fig. 1 are plotted in Fig. 3. Additional data from nearby regions beyond those shown in Fig. 1(k-o) are also incorporated in Fig. 3, enabling a comparison of chemical distributions in similar phases across different locations.

Our observations reveal that post-growth annealing leads to the decomposition/oxidation of $Cu_2S$ and Cu microphases, primarily converting them into $CuO_x$. Post annealing in $O_2$, the majority of $CuO_x$ regions measured have $x$ close to 1, and the O concentration within/near the Pb phase seems changed. It has been established that the O concentration in copper-oxide superconductors correlates with their superconducting properties due to the impact of additional/absent O near Cu on the local electron configuration [23-25]. Whether a similar phenomenon exists in $Pb_{10-x}Cu_x(PO_4)_6O$ remains an open question. Moreover, we observe that the phosphorus elemental concentration in L (P to Pb $\sim 0.24 \pm 0.05$ : 1) is much lower than that in C (P to Pb $\sim 0.64 \pm 0.05$ : 1), and in W (P to Pb $\sim 0.60 \pm 0.08$ : 1). It is important to note that for $Pb_{10}(PO_4)_6O$, the nominal P to Pb atomic ratio is 0.6 : 1. From this, we deduce that the Pb phase in C and W is close to $Pb_{10-x}Cu_x(PO_4)_6O$, while that in L is not the intended LK-99 phase (we note that EDS may not be exactly accurate for the absolute atomic concentrations and our samples seems to have a lower Cu concentration in the Pb phase) [9,10].

To elucidate the microscale magnetism within the samples, dual atomic force (AFM) and diamond nitrogen-vacancy (NV) scans were conducted, using an AFM/NV instrument built on a modified CIQTEK CQDAFM platform. To enable sequential scanning of identical surface areas using both AFM and NV, we integrated an AFM probe with an NV probe array on the same tuning fork within our instrument. SEM images captured prior to the AFM/NV measurements were employed as guiding maps to locate the regions scanned by AFM/NV. Given the presence of phase mixtures, the sample surfaces exhibit a relatively rough texture, displaying a height difference on the order of a few hundred of nanometers among various phases. This height disparity enables the identification of microphase regions. To measure the relatively small magnetic responses of the sample surface, instead of performing a full-range NV frequency scan, we fixed the measurement frequency at 2870 MHz, which corresponds to the energy difference between spin sublevels of the NV ground state [26]. The presence of an even a minute (ferro)magnetic field would cause the resonance frequency to deviate from this value, leading to a change in NV luminescence.

Figure 4 depicts the AFM/NV results obtained from selected surface areas of a C1 sample piece. The NV channel captures local magnetic fields generated by corresponding phases, while the AFM channel visually differentiates distinct phases. To minimize the influence of height difference on the NV measurements, we carefully traced the sample surface using a close-distance noncontact method. By combing AFM, NV, and SEM results (Fig. 4(a-i)), we discover that the micro-size $Cu_2S$ regions can induce a diamagnetic (or, strictly speaking, diamagnetic or an absolute zero-magnetic) local response. In contrast, the Pb phase exhibits a mixture of diamagnetic and weak ferromagnetic responses. Note that the distribution of local magnetism is predominantly influenced by the different phases rather than the local heights. Phase boundaries at the microscale shown in AFM, NV, and SEM results align very well. This observation implies that the volume fraction ratio between Cu-related and Pb phases might significantly affect the overall magnetic attributes of the synthesized samples.

Next, we examined bulk magnetic responses of both the C1 and 99.5%-pure $Cu_2S$ up to 400 K at a magnetic field strength of 0.1 T (as shown in Fig. 4(j,k), $Cu_2S$ mass ~ tens of milligram). Interestingly, the overall magnetic susceptibility of C1 undergoes a sign change at around 8 K, above which a diamagnetic response is observed. A similar magnetic susceptibility behavior is seen for $Cu_2S$, where the sign of susceptibility flips at approximately 5 K. Furthermore, a distinct change in magnetic susceptibility is evident around 370 K for $Cu_2S$, correlating with its structural transition from β to γ phase [18, 27-31]. A small temperature shift in the measurement results is observed after exposure of $Cu_2S$ to air, which causes slight alteration of the transition [28, 32-34]. This change around 370 K is absent in the magnetic susceptibility of the C1 sample, likely due to the relatively small fraction of the $Cu_2S$ phase. To further study possible connections between the Cu-related phase and the diamagnetic response, we located a tiny piece of sample that has the ability to (partially) levitate above permanent magnets. Markedly, it is predominantly composed of $Cu_2S$, with a minor segment of the Pb phase embedded in the center of its surface (as depicted in Fig. 4(l)). We note that typical 99.5%-pure $Cu_2S$ pellets do not levitate in our experiments. A specific combination of mixed phases or the geometry of the piece could be vital in this regard, however, no definitive conclusion is reached in our work.

Microscale chemical phase separations can precipitate dramatic outcomes and intriguing electrical behaviors in superconductors and related materials. For example, percolative superconductor-to-insulator transitions have unveiled an archetypal quantum phase transition, in which charge carriers have to wiggle and tunnel between conductive regions (e.g., [35]). The local electrical properties of our samples were characterized using a Lakeshore CRX-4K probe station and Keysight B2901B precision source/measure units, employing a 4-contact measurement method on the sample surface (Fig. 5(a)). The presence of phase mixtures on the surface can lead to deviations in current flow from the "presumed" path, with regions of low resistance being favored.

Figure 5(b) displays a reversable current switch in local electrical resistivity as a function of the measurement time, when slowly lowering the measurement temperature (near 290 K, ~ -2 K/min). We believe that this type of current path switch can frequently occur in compounds exhibiting microphase separations. To illustrate, Fig. 5(c) compares the electrical resistance measured on an arbitrary long current path on C1, on a $Cu_2S$/Cu dominant phase, and on a Pb dominant phase. A clear jump in electrical resistance of C1 is observable, transitioning from behavior similar to that of $Cu_2S$/Cu to behavior resembling that of the Pb phase. This change signifies a current path switch potentially due to different temperature effects locally. Because the majority of the LK-99 related compounds synthesized so far contain $Cu_2S$ and other Cu-related regions [9-22], these results shed light on the electrical behavior observed in the similar compounds.

Furthermore, we performed current-voltage experiments on 99.5%-pure $Cu_2S$ pellets. Since $Cu_2S$ has a structural transition near 370 to 380 K (dependent on $Cu_2S$ exposure conditions in air), large currents can result in a local temperature variation and thus a structural transition of $Cu_2S$ from a low-resistance state to a high-resistance state [18, 27-33]. In Fig. 5(d), we plotted current-voltage results on a $Cu_2S$ pellet at 348 K by sweeping the electrical current. A resistance transition is observed when the current sweep (heating) time is set to 100 seconds (from 0 to 50 mA) and it is absent when the current sweep (heating) time is limited to 10 seconds (0 to 40 mA). This phenomenon bears resemblance to that reported in AgI, where the resistivity shifts by nearly three orders of magnitude upon transitioning from its α to β structural phase [36]. Whether applying electrical current or heating is the primary cause for the resistance change is unknown.

By characterizing compounds containing $Pb_{10-x}Cu_x(PO_4)_6O$ from three independent groups utilizing comprehensive micro-region techniques, we have determined that the peculiar magnetic and electrical characteristics observed in our compounds can be attributed to phase mixtures at microscales, especially to the presence of $Cu_2S$ and related regions. Despite the extensive past research on the structure of $Cu_2S$ [27-33], details of its electrical and magnetic properties require further studies. Our study underscores the significance of probing physical and chemical properties at the microscale, particularly in the early stages when sizable single crystals are unavailable, as an effective approach to explore novel materials. This research methodology has also proven advantageous in comprehending intricate inhomogeneities and interrelated orders in quantum materials (e.g., [37]).

Acknowledgement:

This work was supported by the National Key Research and Development Program of China (Grants No. 2022YFA1602800, 2022YFA1403400), National Science Foundation of China (Grants No. 12274439, 12134018, 12174424), Chinese Academy of Sciences through the Youth Innovation Promotion Association (2022YSBR-048), and Beijing Natural Science Foundation (Z190008). Y. L. acknowledges support from all who contributed to this project during the supposed "summer leave" at IOP. We thank instrument assistance from Changjiang Zhu at CIQTEK, and Jun Lu and Yuan Li for useful discussions.

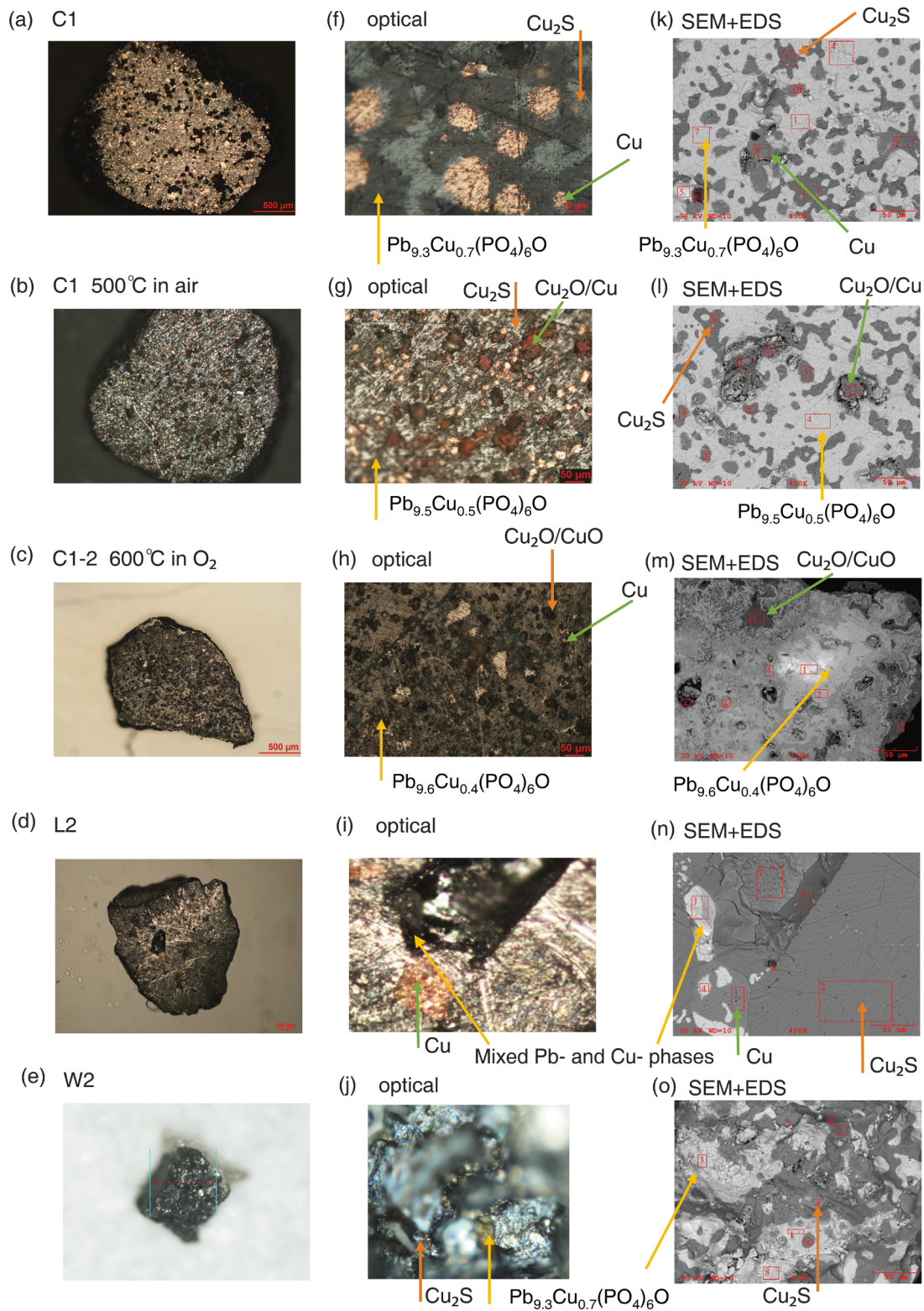

**Fig. 1. Microscale chemical phases of C, L, and W.** Optical microscopy and scanning electron microscopy (SEM) images of samples from three independent synthesis groups. (a-c) Photographic pictures of C1 and C1-2 (a piece broken from C1) in as-grown and post-annealing conditions. (d,e) Photographic pictures of L2 and W2. (f-j) Zoomed-in images showing microscale phase separations of samples in (a-e), respectively. (k-o) SEM back electron scattering images of samples in (a-e). Rectangular boxes indicate areas used to gather energy-dispersive X-ray spectra (EDS) in Fig. 3.
7

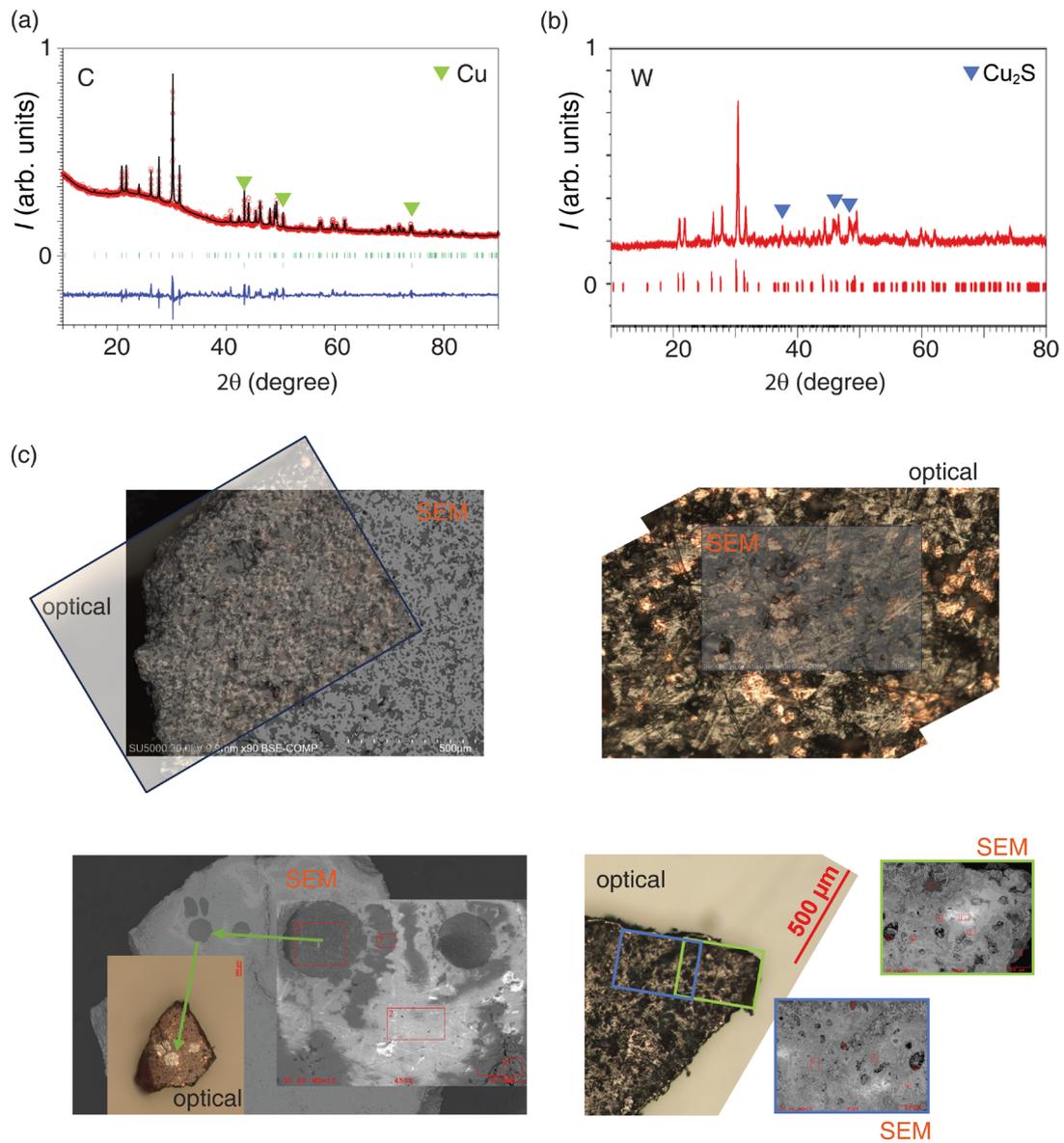

**Fig. 2. X-ray powder diffraction and cross-match of the phases.** (a) $2\theta$-scan of C. Green triangles indicate XRD peaks from Cu. (b) $2\theta$-scan of W. Data taken from ref. [18]. Blue triangles indicate XRD peaks from $Cu_2S$. (c) Demonstrations of phase cross-match process between optical microscopy and SEM images.

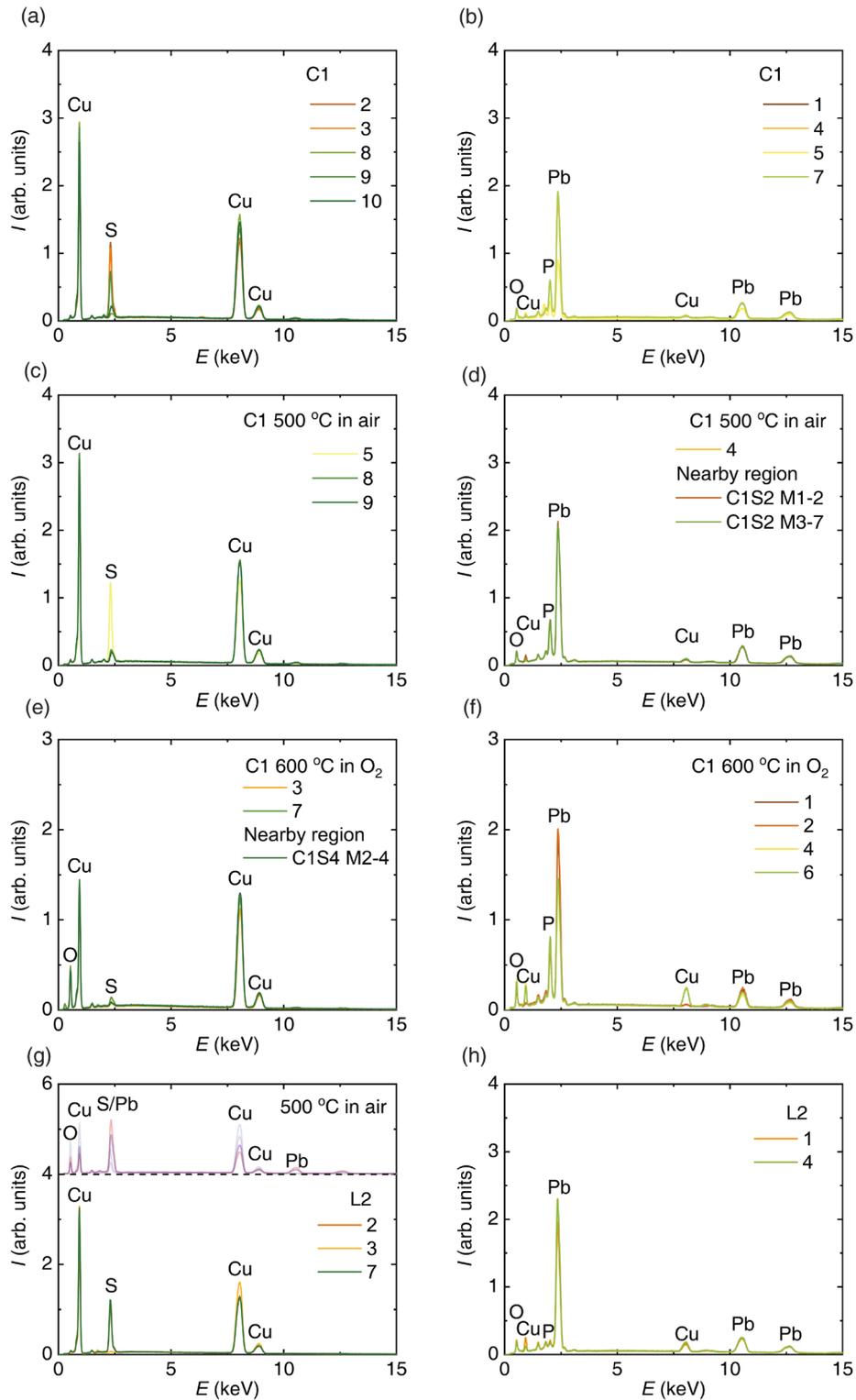

**Fig. 3. Chemical analysis of microscale regions.** Energy-dispersive X-ray spectra (EDS) of samples shown in Fig. 1. (a,b) EDS data of Cu- and Pb-rich regions that correspond to box areas of C1 plotted in Fig. 1(k). (c,d) EDS of Cu- and Pb-rich regions of box areas in Fig. 1(l). Additional EDS data collected from nearby regions (not shown in the image of Fig. 1(l)) are included. (e,f) EDS data corresponding to box areas in Fig. 1(m) with additional data. (g,h) EDS data corresponding to box areas in Fig. 1(n). Extra EDS data after annealing are presented.

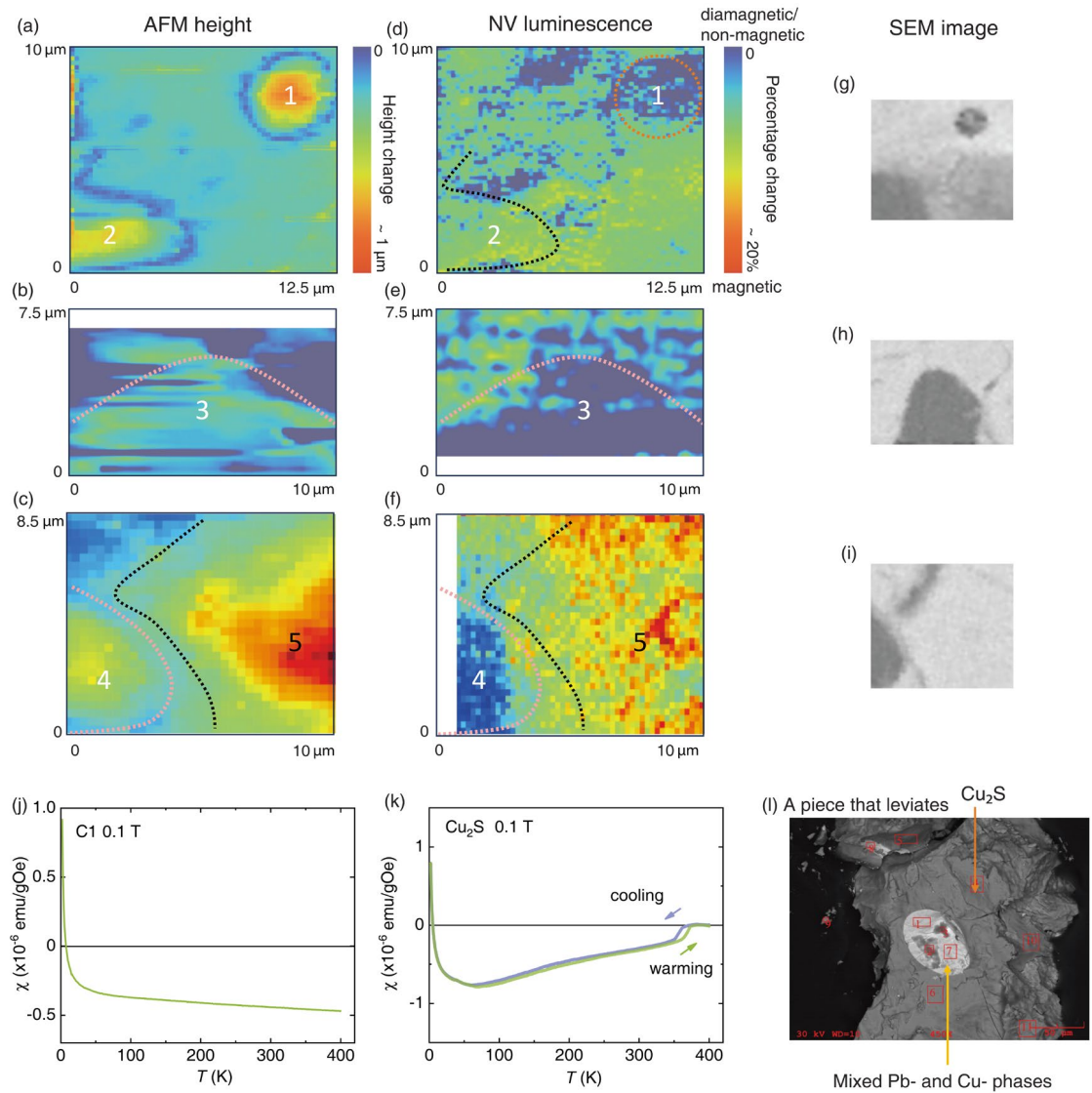

**Fig. 4. Magnetism and chemical phases at the microscale.** (a-c) Representative atomic force microscope (AFM) images for a piece of C1. Distinctive phases have a height difference. (d-f) Diamond nitrogen-vacancy (NV) luminescence acquired by scanning the same areas in Fig. 4(a-c). The NV luminescence reflects the magnetic field generated by various phases. (g-i) SEM region images of C1 that show identified areas measured by AFM and NV in Fig. 4(a-f). (j) Magnetic susceptibility of C1. (k) Magnetic susceptibility of 99.5%-pure $Cu_2S$. (l) SEM image of the sample piece that can (partially) levitate above magnets.

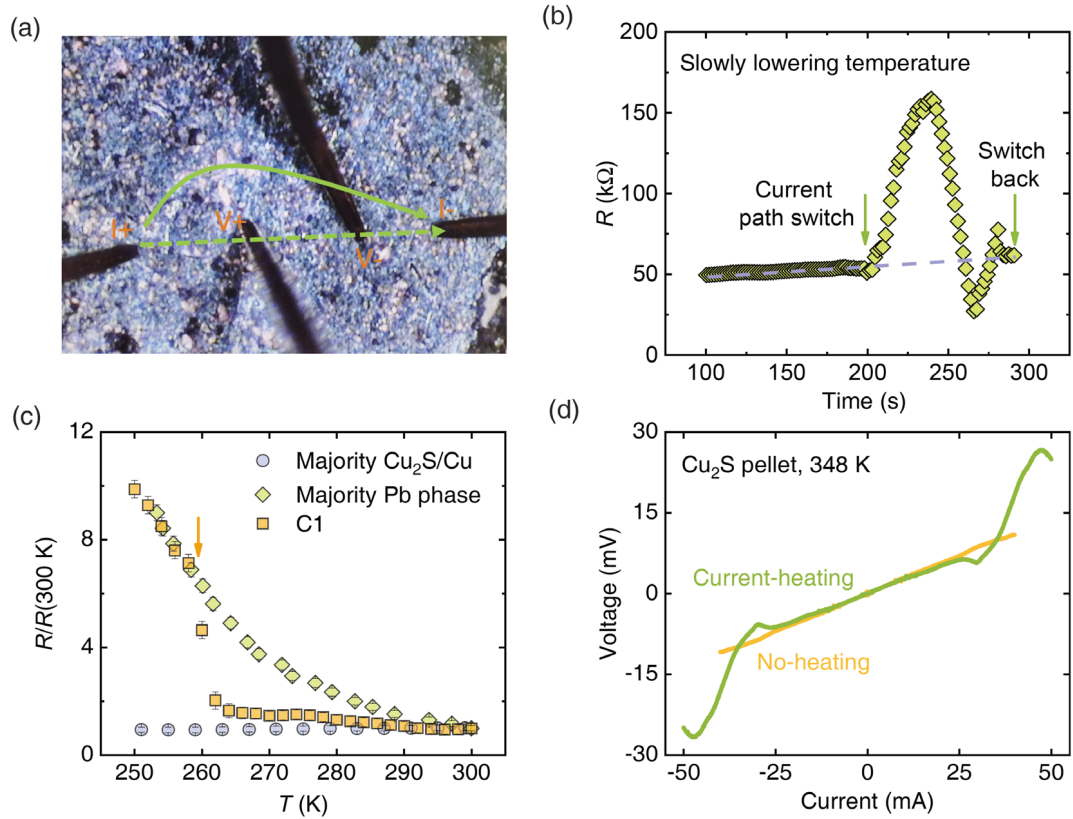

**Fig. 5. Local electrical measurements.** (a) A photographic image depicts the 4-probe electrical measurement on the C1 surface. Dashed and solid curves are illustrations of the "presumed" and a possible actual current path. (b) Time-evolution of surface resistance of a W piece while gradually lowering the measurement temperature. (c) Comparison between the reduced electrical resistance obtained on an arbitrary long path on C1 (orange squares), on a $Cu_2S$/Cu dominant phase (blue circles), and on a Pb dominant phase (green diamonds). (d) Current-voltage measurement for a $Cu_2S$ pellet at 348 K, illustrating a resistance-state transition induced by heating the pellet with electrical current. Positive and negative current sweeps are symmetrized for visualization.

| Samples | Optical image | SEM/EDS | AFM/NV | Electrical resistance | XRD | Broken pieces |
|---|---|---|---|---|---|---|
| C, as-grown | x | x | x | x | x* | ~ 4 |
| C, 400°C in air, 2h | x | | | | | ~ 4 |
| C, 500°C in air, 2h | x | x | x | x | | ~ 4 |
| C, 400°C in $O_2$, 2h | x | | | | | ~ 4 |
| C, 600°C in $O_2$, 8h | x | x | x | x | | ~ 4 |
| L, as-grown | x | x | | | | ~ 6 |
| L, 500°C in air, 2h | x | x | | | | ~ 6 |
| W1, as-grown | x | x | | x | | ~ 2 |
| W2, as-grown | x | x | x | x | x* | ~ 8 |

Table 1. Summary of sample information. $O_2$ annealing was done in a static oxygen environment at ~ 1 atm. *: XRD taken by sample growers.